\documentclass[12pt]{article}
\title{Phase-mixing v.s. phase synchronization in the dynamics of flow-shear induced edge transport barrier}
\author{M. Leconte \\
National Fusion Research Institute, Daejeon 34133, South Korea
}

\usepackage{graphicx}

\newcommand{\dif}{\partial}
\newcommand{\p}{\tilde p}

\newcommand{\chiperp}{\chi_\perp}

\begin{document}
\maketitle

\begin{abstract}
Nonlinear relaxation oscillations of flow-shear induced transport barriers can be qualitatively reproduced using a phenomenological critical-gradient model [M. Leconte, Y.M. Jeon and G.S. Yun, \emph{Contrib. Plasma Phys.} 56, 736 (2016]. Here, we give a more in-depth analysis of the mechanism of these nonlinear oscillations, associated to nonlinear \emph{phase synchronization}, in an extended version of the model including random fluctuations.
\end{abstract}

\section{Introduction}
The effects of $E \times B$ flow shear on transport  is a well-studied phenomenon in fusion plasmas, where it is believed to play an important role in the formation of transport barriers at the plasma edge \cite{ConnorWilson2000, DiamondI2Hahm2005}. Such edge transport barriers are not stationary, but relax quasi-periodically. Several models have been proposed to account for such relaxations \cite{LebedevDiamond1995,Itoh2Fukuyama1996,Itoh2FukuyamaYagi1994, Beyer2005, Constantinescu2011, Diallo2018}. In addition, nonlinear resistive-MHD simulations observed similar nonlinear oscillations \cite{Orain2015}. In previous works \cite{LeconteJeonYun2016, Oh2018}, we proposed a Ginzburg-Landau like phenomenological model, to account for the quasi-periodic dynamics of these relaxations. In the present work, we give a more in-depth analysis of the mechanism of these nonlinear oscillations, associated to nonlinear \emph{phase synchronization}, in an extended version of the model including random fluctuations modeled as white-noise.
We summarize the main results: i) We identify \emph{phase-winding} due to $E \times B$ flow-shear as a possible nonlinear mechanism responsible for transport barrier relaxations, and ii)  There is a critical degree of phase-winding, corresponding to a critical radial-wavenumber above which abrupt un-winding occurs, corresponding to the fast relaxation.
The article is organized as follows: In section 2, we present the model and in section 3, we present the results. In section 4, we discuss the results and give conclusions.

\section{Model}
\begin{figure}
\begin{center}
\includegraphics[width=0.5\linewidth]{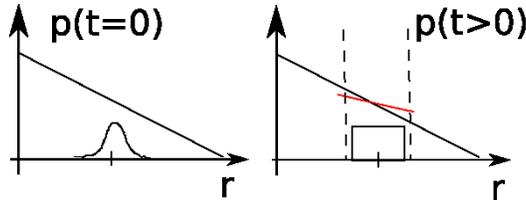}
\caption{Sketch of the critical gradient model: the mode grows at the center of the sheared flow induced barrier, saturates nonlinearly, thus flattening the profile (red-line), and undergoes nonlinear oscillations due to the sheared flow.}
\label{fig1}
\end{center}
\end{figure}

We analyze the following extended Ginzburg-Landau model, a phenomenological model for transport barrier relaxations  \cite{LeconteJeonYun2016,Oh2018}:
%\begin{equation}
%\frac{\dif \p}{\dif t} + i V' L_V \tanh \Big[ \frac{x}{L_V} \Big] \p = \dturb \frac{\dif^2 \p}{\dif x^2} + \gamma_0 \p - \gnl |\p|^2 \p + \tilde \eta
%\end{equation}
%or in normalized form:
\begin{equation}
\frac{\dif \p}{\dif t} + i hV' \tanh \Big[ \frac{x}{h} \Big] \p = \frac{\dif^2 \p}{\dif x^2} + (Q - Q_c) \p - |\p|^2 \p + \tilde \eta
\label{eq1}
\end{equation}
where time is normalized to the inverse growth-rate $\gamma_0^{-1}$, and space is normalized to the coherence length $\xi = \sqrt{\chi_\perp / \gamma_0}$, with $\chi_\perp$ the heat diffusivity, assumed constant in this model. The flow shear magnitude $V'$ is normalized as $k_y V' \xi / \gamma_0 \to V'$, and it has maximal shear at $x=0$ (center of the shear-layer) and is exponentially-decaying away from $x=0$, with a shear-layer width $h$. Here, $x$ and $y$ denote the local radial and poloidal directions in a fusion device. The control parameter $Q$ denotes the total heat flux, and $Q_c$ is the linear threshold, above which the mode grows at the center of the transport barrier.
The model includes Gaussian white-noise, with correlation $\langle \eta(x,t) \eta(x',t') \rangle = \eta_0 \delta(x-x') \delta(t-t')$, and $\eta_0 = 5$. \\
In Eq. (\ref{eq1}), $\tilde p$ represents the nonlinearly-modulated amplitude of a Fourier mode, i.e. $\delta p(x,y,t) = \tilde p(x,t) e^{i k_y y} +c.c.$ This mode grows at the center of the sheared-flow induced transport barrier at the plasma edge, saturates by flattening the profile and later its amplitude oscillates quasi-periodically due to the sheared flow, thus causing quasi-periodic relaxations of the transport barrier. This mechanism is sketched [Fig. \ref{fig1}].
We now justify and describe the physics present in this phenomenological model. The second term on the l.h.s. of Eq. (\ref{eq1}) represents the advection by the mean $E \times B$ sheared flow. The first term on the r.h.s. is a diffusion term due to residual microturbulence. The second term on the r.h.s. is the driving term, which corresponds to energy gain from the unperturbed mean pressure gradient, above the threshold $Q \ge Q_c$. The third term on the r.h.s. represents the effect of the reduction of the mean pressure gradient due to the convective heat flux $Q_{\rm conv}  = \sum_k \tilde p^* \tilde v_{Ex} +c.c$, where $\tilde v_{Ex} = -i k_y \tilde \phi / B$ denotes the radial $E \times B$ velocity, with $\tilde \phi$ the electric potential. For a single coherent mode, as we consider here, the convective flux reduces to: $Q_{\rm conv} \simeq \tilde p^* \tilde v_{Ex} +c.c$ . As we consider an interchange type of instability, we assume that the electric potential $\tilde \phi$ follows the pressure $\tilde p$ with a $\pi / 2$ phase delay, corresponding to maximal convective transport, i.e. $\tilde \phi \sim i \tilde p$, or equivalently $\tilde v_{Ex} \sim \tilde p$. Hence, in this model, the convective heat flux is quadratic in the mode amplitude $Q_{conv} \propto |\tilde p|^2$. Then, we employ heat balance to relate the total mean pressure gradient $\dif_x \langle p \rangle$ to the mode amplitude, where $\langle \cdot \rangle = \int \cdot dy$ denotes a flux-surface average. Heat balance at steady-state reads: $Q_{conv} + Q_{coll} = Q$, where $Q = {\rm Cst}$ is the total heat flux, related to the heating power, and $Q_{coll} = - \chiperp \dif_x \langle p \rangle$ is the diffusive flux, due to collisions and residual microturbulence. Using the expression for the convective flux yields the desired relation for the pressure gradient: $\dif_x \langle p \rangle = p_{eq}' + \dif_x \Delta p_{eq}$, with $p_{eq}' \propto - Q <0$ the unperturbed pressure gradient, and $\Delta p_{eq} \propto |\tilde p|^2 >0$ the back-reaction of the mode on the pressure gradient. Hence, the back-reaction of the mode on the pressure profile is modeled as  a self-damping term $- |\tilde p^2| \tilde p$.

The parameters used for the numerical calculations are: $V'=4 \times 10^1$, $h=1  \times 10^{-2}$, $Q=10$ and $Q_c = 0.5$, and $\chiperp=1 \times 10^{-2}$.
In a previous publication \cite{Oh2018}, sensitivity studies were undertaken to determine the sensitivity of the model to the value of several parameters of the model, namely the flow shear $V'$, the shear-layer width $h$, the distance to threshold $Q - Q_c$ and the heat diffusivity $\chi_\perp$. Depending on the values of $Q - Q_c$ and $\chi_\perp$ (denoted $\gamma_L$ and $\eta$, c.f. Fig 1 in this reference), the equilibrium state $\tilde p =0$ was shown to be stable or unstable. For each value of the cross-field diffusivity $\chi_\perp$, there exists an instability threshold $Q > Q_{th}(\chi_\perp)$ above which the mode is linearly unstable, as expected.  In addition, depending on the values of the flow shear $V'$ and the inverse of the shear-layer width (denoted respectively $A$ and $K$, c.f. Fig. 2 of this reference), it was found that the numerical solution either converges to the stationary saturated state $|\tilde p|  = {\rm Cst} > 0$, or to a state of nonlinear oscillations, and for a given value of the shear-layer width $h$, there is a critical value of the flow shear $V' > V'_c(h)$ above which nonlinear oscillations set in. Note that for small values of the shear-layer width $h$ (large values of $K=1/h$), this critical flow-shear becomes independent of $h$.

\section{Results}

\subsection{Phase-mixing}
\begin{figure}
\includegraphics[width=0.5\linewidth]{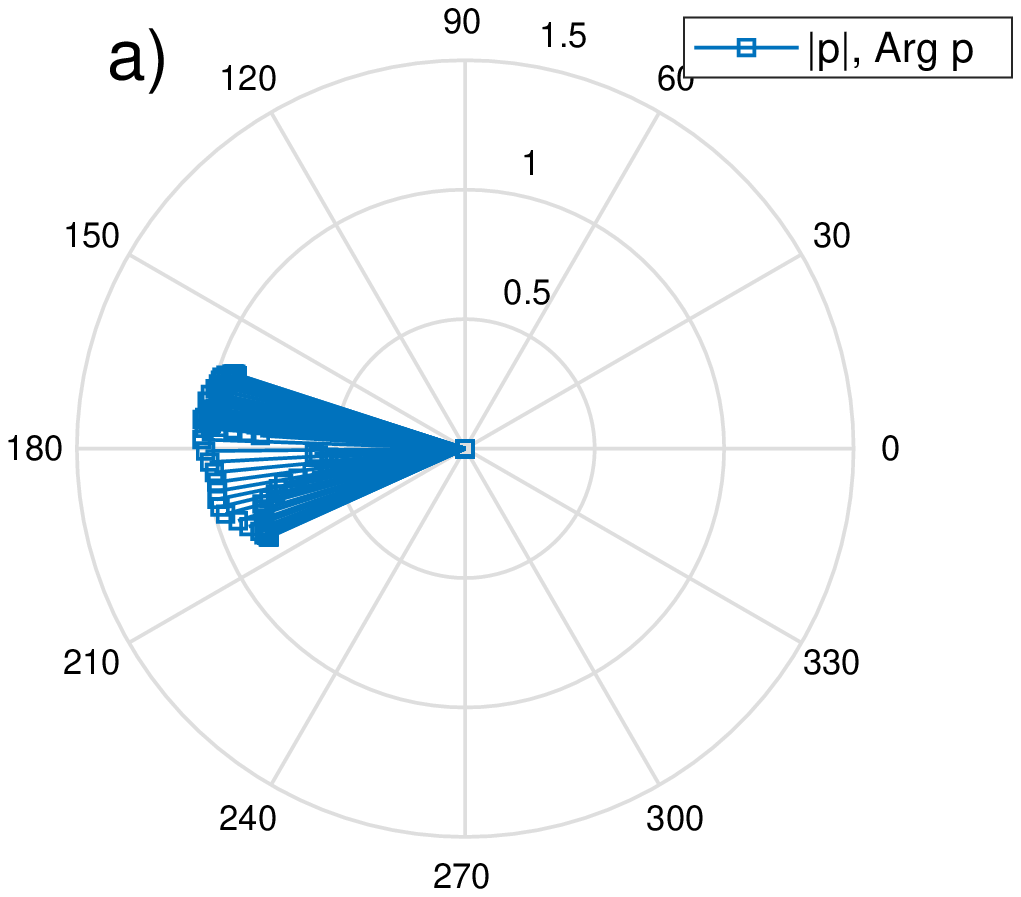}\includegraphics[width=0.5\linewidth]{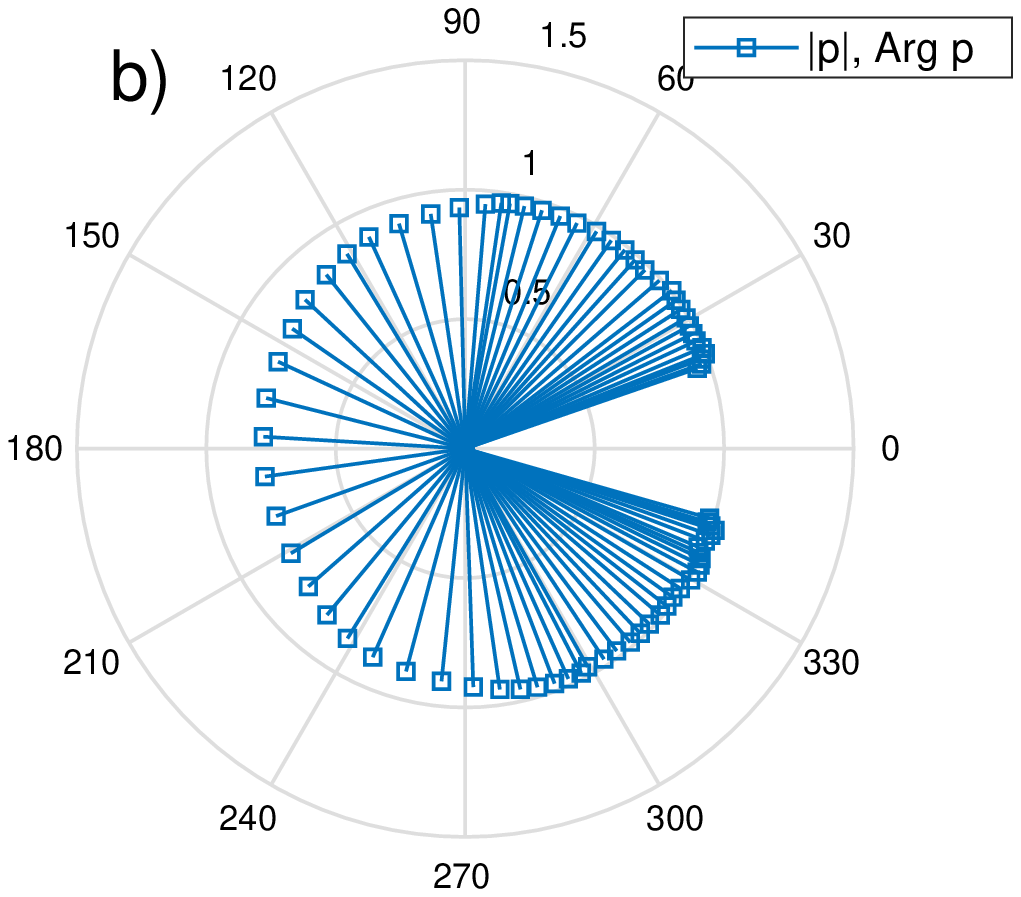}
\caption{Phase-mixing induced by the sheared flow. Polar plot of the amplitude $|\p|$, and phase ($\arg \p$). (a) before phase-mixing, and (b) after phase-mixing. The squares represent the distribution of amplitudes $|\p|$, and the angle of the rods represents the associated ditribution of phases. }
\label{ph-mix}
\end{figure}

The phases are initially random, and subsequently cooperate to form a dissipative structure, with uniform amplitude. Once the saturated state has been reached, the flow shear starts to have an effect [Fig. \ref{ph-mix}a]. Note that the formation of the saturated state is exponentially fast, so the system is already in a saturated-state, before the flow-shear starts to have an effect. The flow-shear induces phase mixing [Fig. \ref{ph-mix}b].  The phase-mixing effect due to the flow-shear is responsible for a scattering of the phases, where at every radial location, the phase $\Theta(x,t)$ is Doppler-shifted by the local $E \times B$ frequency $\omega(x) = \omega_E(x)$. In this early stage, nonlinear effects on the phase are negligeable, and the mode amplitude decreases in time due to the phase-mixing [Fig. \ref{ph-mix}b].

\subsection{Phase-synchronization}
\begin{figure}
\includegraphics[width=0.5\linewidth]{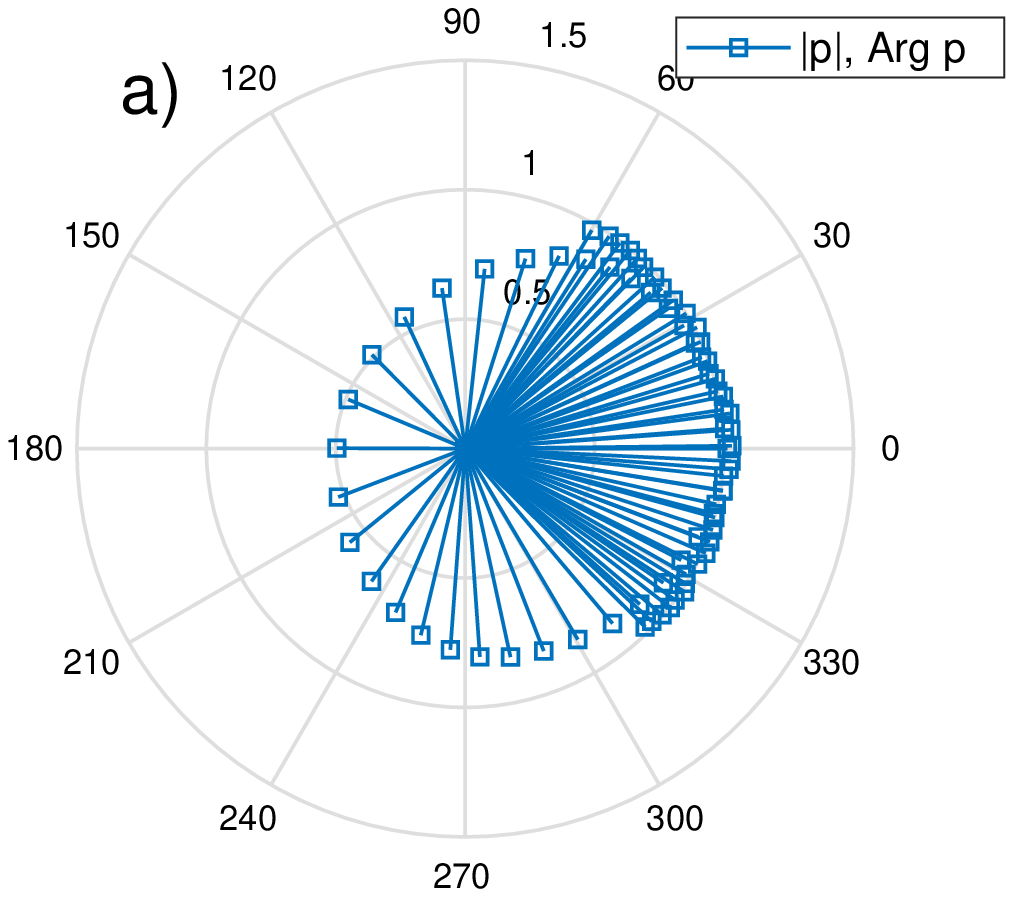}\includegraphics[width=0.5\linewidth]{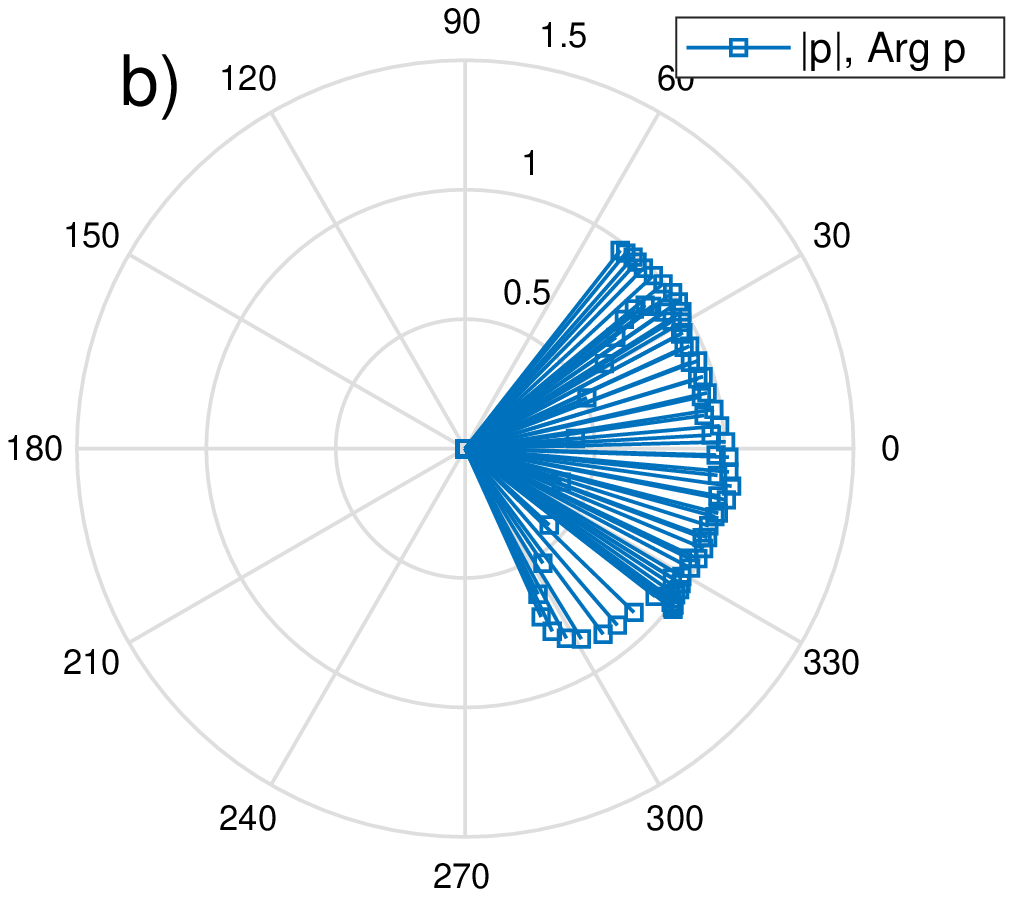}
\caption{Phase-synchronization due to the nonlinearity. Phase distribution: (a) before phase synchronization and (b) after phase synchronization.}
\label{ph-sync}
\end{figure}

\begin{figure}
\includegraphics[width=0.5\linewidth]{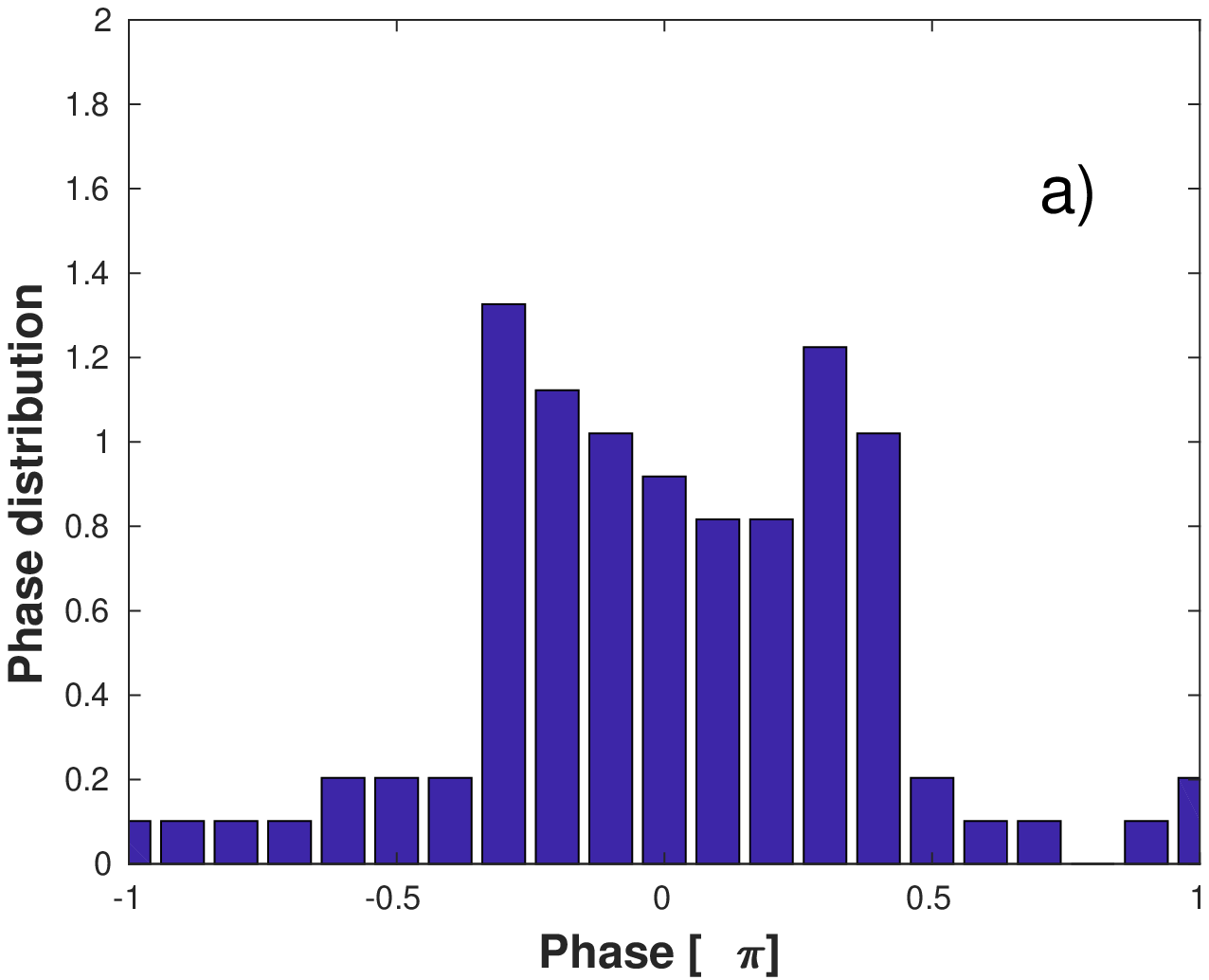}\includegraphics[width=0.5\linewidth]{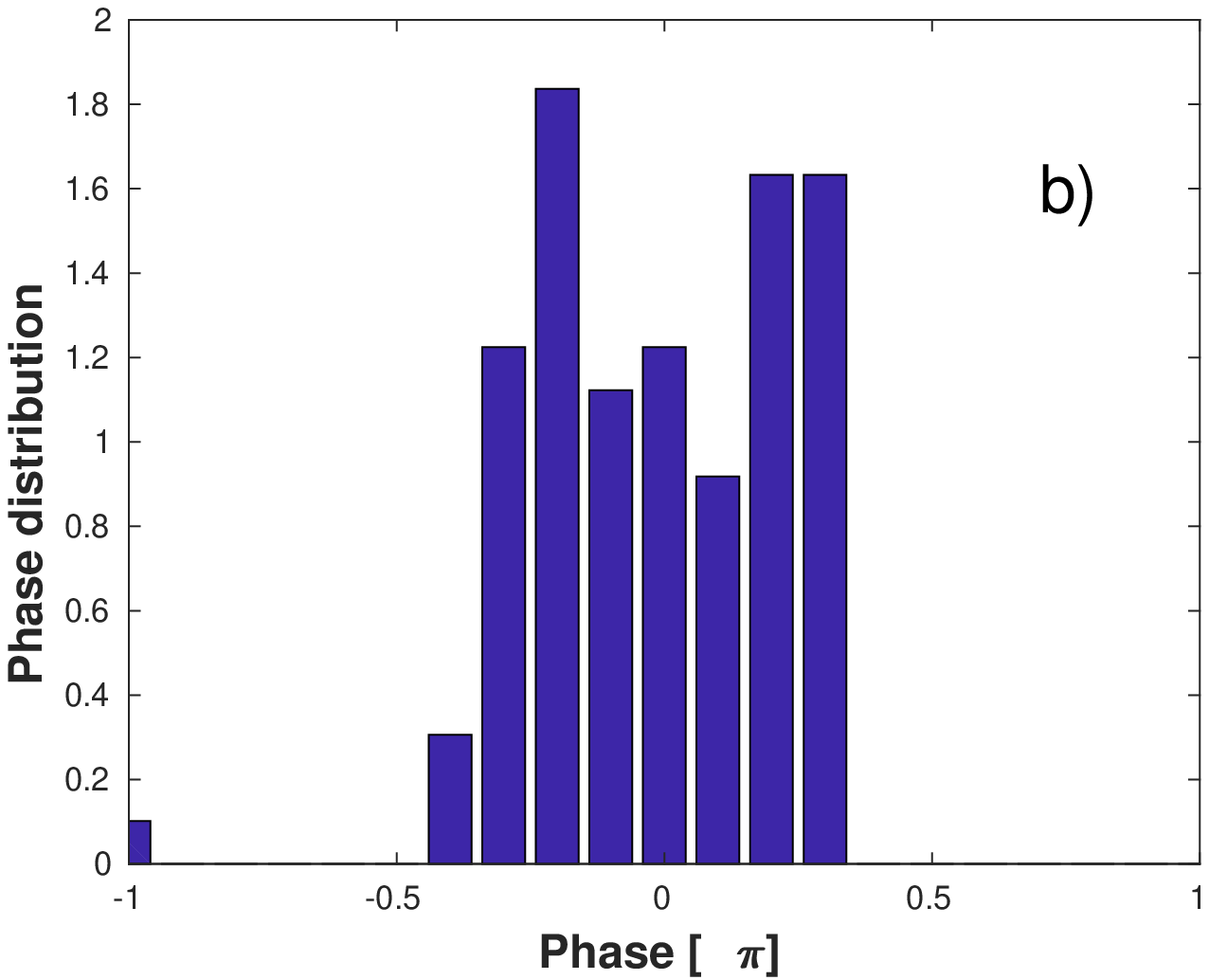}
\caption{Phase distribution: (a) before and (b) after phase synchronization.}
\label{phdist-sync}
\end{figure}

Later on, as the phase gradient, i.e. the radial wavenumber $k_r = \dif_r \Theta$, increases due to the phase-mixing, nonlinear effects start to become important. Note that the magnitude of the radial wavenumber corresponds to the degree of winding of the phases on the phase distribution diagram [Fig. \ref{ph-sync}a]. When this winding exceeds a threshold, nonlinear effects set in and phase synchronization occurs, which corresponds to the abrupt 'un-winding' of the phases in the phase distribution diagram [Fig. \ref{ph-sync}b]. This synchronization phenomenon is also shown on the histogram of phases, the phase distribution diagram [Fig. \ref{phdist-sync}]. Before synchronization [Fig. \ref{phdist-sync}a] the system shows a rather broad distribution of phases, consistent with the phase-mixing effect, while after synchronization [Fig. \ref{phdist-sync}b], the phase distribution becomes peaked around $\Theta \simeq 0$.

\begin{figure}
\begin{center}
\includegraphics[width=0.5\linewidth]{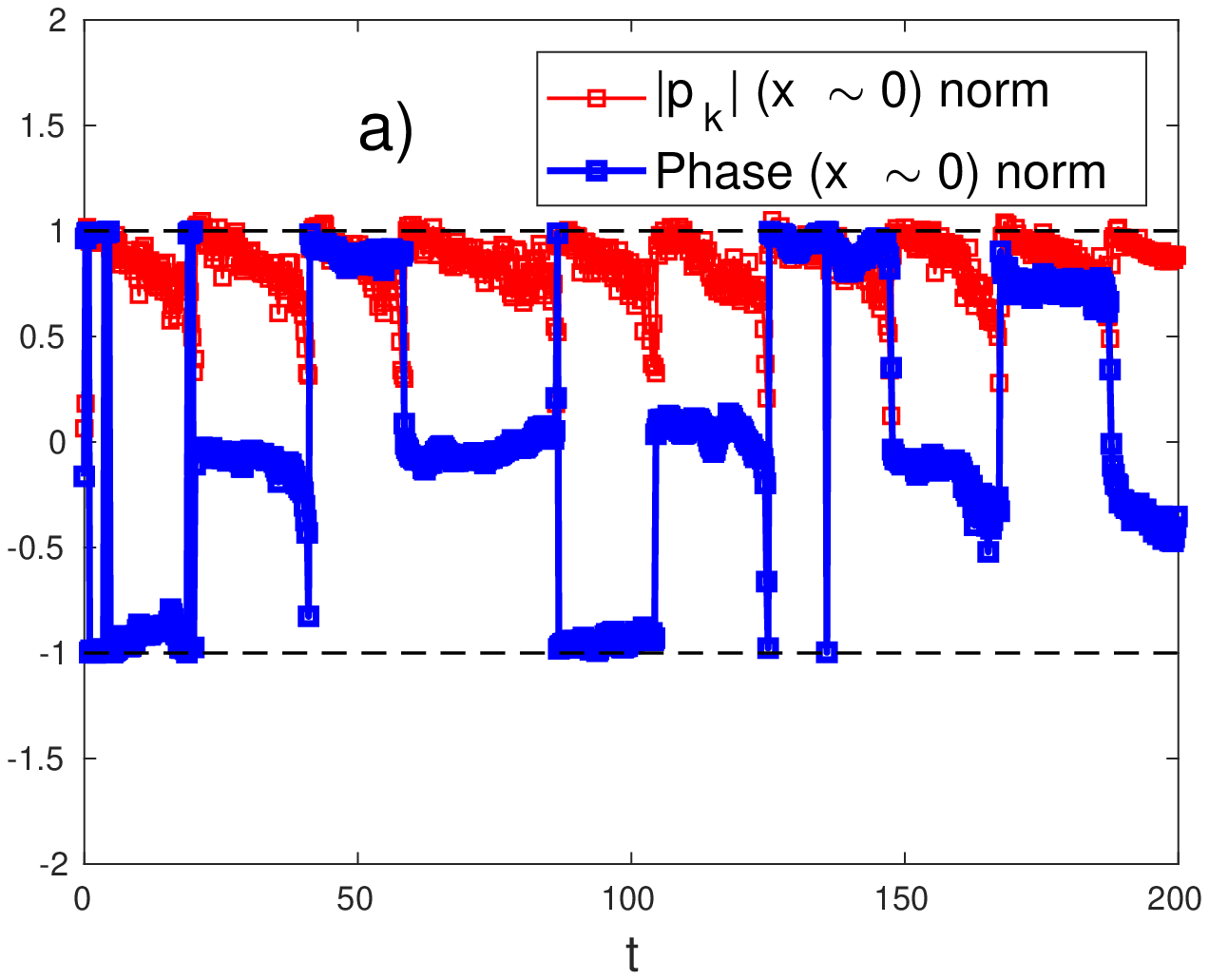}\includegraphics[width=0.5\linewidth]{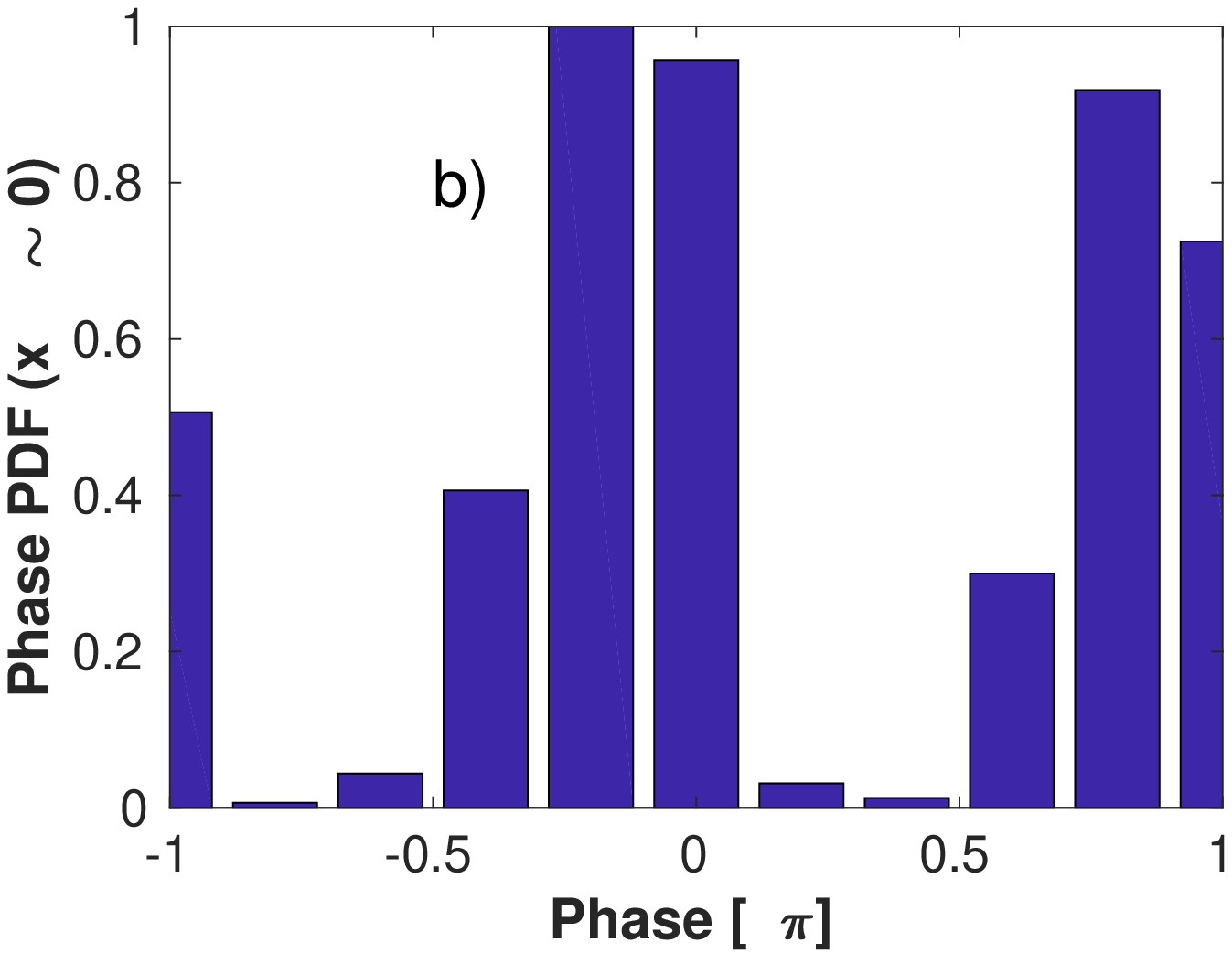}
\caption{(a) Time evolution of amplitude $\p$ (red open squares) and phase $\Theta = \arg \p$ (blue filled squares) at the barrier center $x=0$. (b) Probability density function of the phase.}
\label{fig-histtime}
\end{center}
\end{figure}

Next, we plot the time series of the signal amplitude $|\p|$ (red) and the associated phase $\Theta = \arg \p$ (blue), at the center of the barrier [\ref{fig-histtime}a]. The amplitude exhibits nonlinear oscillations, while the phase jumps from $\Theta \simeq 0$ to $\Theta \simeq \pi$. The associated probability distribution (PDF) of the phase is shown [Fig. \ref{fig-histtime}b]. It shows two peaks, one at $\Theta \simeq 0$, and one at $\Theta \simeq \pi$. Note that there is an additional peak at $\Theta = -\pi$ but because the phases are $2 \pi$-periodic, a value of $- \pi$ is equivalent to $\pi$, so this additional peak really belongs to the peak at $\Theta \sim \pi$.
Note that the dynamics displays sensitivity to initial conditions. If initialized with a given phase distributed around $\frac{\pi}{2}$, instead of a random phase, the phase jumps from $\Theta \simeq \frac{\pi}{2}$ to $\Theta \simeq - \frac{\pi}{2}$, instead of from $\Theta \simeq 0$ to $\Theta \simeq \pi$ (not shown).

\begin{figure}
\begin{center}
\includegraphics[width=0.5\linewidth]{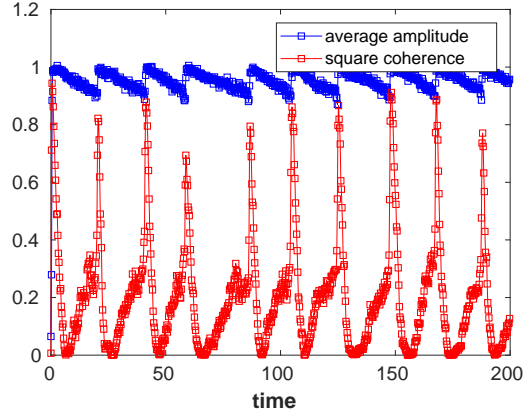}
\caption{Phase coherence squared (red), and radially-averaged amplitude (blue) v.s. time.}
\label{fig-phcoh}
\end{center}
\end{figure}

The phase coherence $R(t) = \frac{1}{N_x} \sum e^{i \Theta_j}$ is a measure of the degree of phase synchronization. Here, $N_x$ is the total number of phases (the number of points in the radial direction $x$), $\Theta_j = \Theta(j \Delta x, t)$ are the phases. and $j=1,\ldots, N_x$ is their index. We show the phase coherence squared $R^2(t)$ v.s. time [Fig. \ref{fig-phcoh}]. Each barrier relaxation - i.e. sudden increase in the mode amplitude - is synchronous with a peak in the square coherence.

\begin{figure}
\begin{center}
\includegraphics[width=0.5\linewidth]{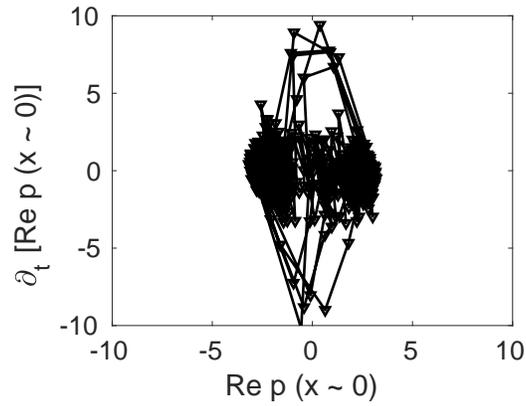}
\caption{Representation of the signal in dynamical phase-space ($\Re (\tilde p), \Re (\dif_t \tilde p)$). It shows a limit-cycle elongated along the y-axis, a typical feature of relaxation oscillations.}
\label{fig-phsp}
\end{center}
\end{figure}

The dynamics of the model resembles limit-cycle oscillations. To show this, we plot the signal in dynamical phase-space [Fig \ref{fig-phsp}]. The two axis of the plot are the signal itself (${\rm Re} ~\tilde p$) on the x-axis, and the time derivative of the signal (${\rm Re} ~\dif_t \tilde p$) on the y-axis.

\section{Discussion and conclusions}
First, we wish to stress that the 'phase' refered to in this article, is \emph{not} the transport crossphase between the pressure and potential fields which was investigated in Ref. \cite{XiXuDiamond2014} for peeling-ballooning modes. In our model, the transport crossphase is assumed not to be affected by the equilibrium flow shear (due to Galilean invariance) and is set so as to maximize the convective heat transport $Q_{conv} \simeq |\tilde p|^2 \sin \delta$ (i.e. it is set to $\delta = \pi / 2$). The 'phase' that we refer to is really the eikonal familiar from the WKB formalism. In other words, the phase $\Theta$ is defined such that its radial derivative is the radial wavenumber $k_r = \dif_r \Theta$ and its time-derivative is minus the instantaneous frequency $\omega = - \dif_t \Theta$.
The phase synchronization phenomenon described in this work is similar to phase synchronization in a group of many coupled phase oscillators, i.e. the Kuramoto model \cite{Acebron2005}. We may link the phase synchronization phenomenon observed on Figs. (\ref{ph-sync}b) and (\ref{phdist-sync}b) to the frequency-clustering, i.e. the formation of frequency plateaus \cite{KirnerRossler1997}. Each jump from $\Theta = 0$ to $\Theta = \pm \pi$ in the timeseries of the phase [Fig \ref{fig-histtime}] is also associated to this nonlinear mechanism of phase synchronization. We identify \emph{phase-winding} due to $E \times B$ flow-shear, as shown on [Fig. \ref{ph-sync}a] as the nonlinear mechanism responsible for the transport barrier relaxations. There is a critical degree of phase-winding, corresponding to a critical radial-wavenumber above which abrupt un-winding occurs, corresponding to the fast relaxation. This means that the system is of the fast-slow type,  there are two time-scales, a slow timescale corresponding to the slow winding of the phases, and a fast timescale corresponding to the fast un-winding. The presence of slow and fast timescale is also present in the well-known Vanderpol oscillator model.

There are limitations to our model. As the model is phenomenological in nature, it has free parameters: the flow shear magnitude, the shear-layer width, the heating power and the critical heating power.
Another approximation we make is related to the structure of the model itself. As is well-known, pressure-gradient driven instabilities are generally described by an equation second-order in time, with an unstable and a stable branch \cite{Constantinescu2011}, whereas our model is first-order in time. This is justified provided the stable branch has negligeable influence on the dynamics of the model, a reasonable assumption.
In conclusion, we have shown the connection between relaxation oscillations and phase synchronization in a phenomenological model of transport barrier relaxations.

\section*{Acknowledgements}
The author would like to thank G.S. Yun, A. Diallo and P.H. Diamond for usefull discussions, and D.H. Park for pointing to the measure of phase coherence.
This work was supported by R\&D Program through National Fusion Research Institute (NFRI) funded by the Ministry of Science and ICT of the Republic of Korea (No. NFRI-EN1941-5).

\end{document}